\documentclass[preprint,showpacs,preprintnumbers,amsmath,amssymb]{revtex4}

\usepackage{graphicx}
\usepackage{dcolumn}
\usepackage{bm}

\begin{document}



\title{Search for optimal measure for discriminating spike trains with different randomness}

\author{Keiji Miura}
\email{miura@ton.scphys.kyoto-u.ac.jp}
\affiliation{Department of Physics, Graduate School of Sciences, Kyoto University Kyoto 606-8502, Japan}
\affiliation{``Intelligent Cooperation and Control'', PRESTO, JST, c/o The University of Tokyo, Chiba 277--8561, Japan\\}

\author{Masato Okada}
\email{okada@k.u-tokyo.ac.jp}
\affiliation{Department of Complexity Science and Engineering, Graduate School of Frontier Sciences, The University of Tokyo, Chiba 277-8561, Japan}
\affiliation{``Intelligent Cooperation and Control'', PRESTO, JST, c/o The University of Tokyo, Chiba 277-8561, Japan\\}
\affiliation{Laboratory for Mathematical Neuroscience, RIKEN Brain Science Institute, Saitama 351-0198, Japan}

\author{Shigeru Shinomoto}
\email{shinomoto@scphys.kyoto-u.ac.jp}
\affiliation{Department of Physics, Graduate School of Sciences, Kyoto University Kyoto 606-8502, Japan}

\date{\today}

\begin{abstract}
We wish to discriminate spike sequences based on the degree of irregularity. For this 
purpose, we search for a rational expressions of quadratic functions of 
consecutive interspike intervals that efficiently measures 
spiking irregularity. Under natural assumptions, the functional form of the coefficient 
can be parameterized by a single parameter. The parameter is determined so as to 
maximize the mutual information between the distributions of coefficients computed for 
spike sequences derived from different renewal point processes. We find that the local 
variation of interspike intervals, $L_V$ (Neural Comput. Vol. 15, pp. 2823-42, 2003), is 
nearly optimal for whose intrinsic irregularity is close to that of experimental data.
\end{abstract}

\pacs{Valid PACS appear here}

\maketitle

\section{\label{sec1}Introduction}

It is important to extract as much information as possible from spike sequences when 
looking for correlations between animal behaviors and neuronal activities 
\cite{georgopoulos,miyashita,funahashi,fujita} or controlling prosthetic apparatuses by 
neuronal activities \cite{chapin}. In many cases, however, only the mean firing rate is
considered and the timing information is not taken into account. Consideration of detailed
temporal structure of each spike train would help to decode brain signals more efficiently.
We would like to propose a measure, which augments the information provided by the mean
firing rate.

Coefficients that are functions of the interspike intervals (ISIs) are effective 
in detecting a spiking irregularity from a short spike train. For instance, the coefficient 
of variation, $C_V$, is widely adopted as a measure of the variance of ISIs
\cite{cox,abbott,shinomoto1,shinomoto4}.
Recently, a measure of the local variation of interspike intervals, 
$L_V$, was proposed \cite{shinomoto}, as a natural extension of 
$C_{V2}$ which was designed to detect a stepwise variation of consecutive ISIs 
\cite{holt}.  An analysis using $L_V$ revealed that \textit{in vivo} spike sequences are 
not uniformly random, but possess specific characteristics that vary among individual 
neurons. In addition, it was found that the neocortex consists of heterogeneous neurons 
that differ not only from one cortical area to another, but also from one layer to another 
in their spiking patterns \cite{shinomoto7}.

In the present study, we try to modify $L_V$ in an attempt to find a better measure for 
discriminating spike sequences based on the degree of irregularity. Namely, we examine 
rational expressions of quadratic functions of consecutive interspike intervals for 
suitability as coefficients for measuring spiking irregularity. Under reasonable 
assumptions, the functional form of the coefficient is found to be parameterized by a 
single parameter. The parameter is determined so as to maximize the mutual 
information between the distributions of coefficients computed for finite size sample 
sequences derived from different renewal gamma processes. It is found that $L_V$ is 
not optimal for nearly random Poisson spike trains but optimal for more regular spike 
trains.

In Sec.~\ref{sec2}, we explain how we generated spike sequences with the same 
firing rate but different intrinsic irregularity. We show that a gamma 
distribution suffices for that purpose and that two parameters in the gamma 
distribution can be chosen as orthogonal coordinates.
In Sec.~\ref{sec3} we explain $L_V$ and compare it with $C_V$.
We show that attractiveness of $L_V$ stems from its symmetries.
In Sec.~\ref{sec4} we extend $L_V$ and show that, under reasonable assumptions,
the extension of $L_V$ can be parameterized by a single parameter.
In Sec.~\ref{sec5} we explain how we determined the optimal value of the 
parameter using the maximization principle of mutual information.
In Sec.~\ref{sec6}, we determine the optimal value numerically.
In Sec.~\ref{sec7}, we describe our theory, developed using a Gaussian
approximation, for explaining the results.
In Sec.~\ref{TD}, we discuss two non-stationary cases.

\section{\label{sec2}Generating spike trains with different randomness}
In this section, we explain how to generate spike trains with the same firing
rate but different randomness.

There are many ways to generate spike trains artificially.
For example, we can generate spike trains by using a network of spiking neuron
models.
However, we do not need to describe precise spike timing here, and 
a simple mechanism is desirable.
Therefore, we assume that the mechanism is a renewal process and 
that the inter spike interval (ISI) follows a gamma distribution \cite{cox}, 
which is described as
\begin{equation}
p(T) = \frac{1}{\Gamma(\kappa)}\left(\frac{\kappa}{\mu}\right)^\kappa T^{\kappa-1}e^{-\frac{\kappa}{\mu} T},
\label{gamma}
\end{equation}
where $T$ denotes an ISI.
We generate ISIs from the distribution and align them to make a spike train.
The mean and variance of the ISIs are
\begin{equation}
\left\{ \begin{array}{c}
Ex(T)=\mu\\
Var(T)=\frac{\mu^2}{\kappa}.
\label{expectation}
\end{array}\right.
\end{equation}
The mean firing rate is obtained by taking the inverse of the mean ISI 
\cite{lansky}.
The $\kappa$ is a shape parameter; $\kappa=1$ corresponds to an exponential
distribution, and, as $\kappa$ increases, the distribution approaches a normal
distribution.
The exponential distribution corresponds to a Poisson process in which the
firing rate (hazard function) is constant with time independent of the
previous firing time. The spike train looks random.
As $\kappa$ increases, the variance of the ISIs decreases, and the ISIs become
regular.

Our goal is to find an optimal measure for discriminating two spike trains
with different randomness independent of their mean firing rates.
A gamma distribution is suitable for that purpose.
First, we can control the mean firing rate and randomness independently by 
changing the two parameters ($\mu$ and $\kappa$) in the distribution.
Next, experimental data can be well fitted by the distribution.
For example, Baker et al. showed that the spike patterns recorded from 
primary and supplementary motor areas are explicable using a gamma
distribution \cite{baker}.

We can transform the parameters in a gamma distribution arbitrarily.
For example, we can transform the parameters into $(\alpha,\lambda)$:
\begin{equation}
\left\{ \begin{array}{c}
\alpha = \kappa,\\
\lambda = \frac{\kappa}{\mu}.
\end{array}\right.
\end{equation}
The gamma distribution in this coordinate can be written as
\begin{equation}
p(T) = \frac{\lambda^\alpha}{\Gamma(\alpha)}T^{\alpha-1}e^{-\lambda T},
\end{equation}
where $\lambda$ is a scale parameter.
The mean and variance of the ISIs can be written as functions of $\alpha$ 
and $\lambda$ as
\begin{equation}
\left\{ \begin{array}{c}
Ex(T)=\frac{\alpha}{\lambda},\\
Var(T)=\frac{\alpha}{\lambda^2}.
\end{array}\right.
\end{equation}
Thus, there are many ways of writing (parameterizing) a gamma distribution.
We used the expression shown as Eq.~(\ref{gamma}) because $\mu$ corresponds to
the mean ISI and $\kappa$ is orthogonal to it in the sense of information
geometry \cite{amari2,amari3}.
The proof is shown in APPENDIX \ref{appendixA}.
We call the parameters of a gamma distribution coordinates because we regard
the family of gamma distributions as manifold.
We would like to define randomness as information orthogonal to the firing
rate.
Therefore, we regard $\kappa$ as randomness in what follows.
We generate spike trains having different intrinsic randomness by using the
gamma distributions with different values of $\kappa$.

\section{\label{sec3}$L_V$ and $C_V$}

\begin{figure}[t]
\includegraphics[width=70mm]{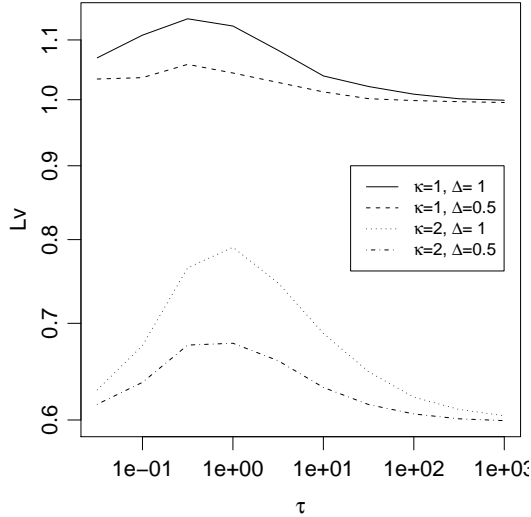}
 \caption{\label{TDGlv}$L_V$ for doubly stochastic gamma process with various
values of time constant $\tau$ and rate amplitude $\Delta$.}
\end{figure}

The measure of local variation proposed by Shinomoto et al. \cite{shinomoto}
is defined as
\begin{equation}
L_V = \frac{1}{n-1}\sum_{i=1}^{n-1} \frac{3(T_i-T_{i+1})^2}{(T_i+T_{i+1})^2},
\end{equation}
where $T_i$ denotes the i-th ISI in a spike train.
The coefficient ``3'' is multiplied so that $\overline{L_V}$ is 1 for a Poisson
process.
$L_V$ is large when consecutive ISIs differ.
It is dimensionless and invariant if all the ISIs are multiplied by a constant.
The conventional Cv is defined as \cite{holt}
\begin{equation}
C_V \equiv \frac{\sqrt{Var(T)}}{Ex(T)}.
\end{equation}
Next we examine the difference between $L_V$ and $C_V$ and calculate $L_V$ and
$C_V$ for the rate modulated gamma process.

We define a rate modulated gamma process as an extension of a gamma 
distribution where the firing rate, $\lambda(t)(=\frac{1}{\mu(t)})$, is 
time-dependent while $\kappa$ is time-independent.
The spikes for the rate modulated gamma process are generated as follows
\cite{abbott,brown}.
Note that we consider only the case of integer $\kappa$.
A spike is generated with probability $\lambda(t) dt$ for every small time 
step, dt.
To be precise, we generate a uniform random number and if it is less than
$\lambda(t) dt$, we generate a spike at that time step.
For the case where $\kappa$ is larger than 1, we keep every $\kappa$-th spike
and remove the others. What is left is the desired sequence.
In fact, for the case where $\lambda$ is constant over time, the spike
sequence generated in this way is equivalent to that generated from a renewal
gamma distribution with $\mu=\frac{1}{\lambda}$.

Here we consider a doubly stochastic gamma process whose firing rate obeys the
Ornstein-Uhlenbeck process \cite{shinomoto6}.
We assume the firing rate, $\lambda$, satisfies
\begin{equation}
\frac{d\lambda}{dt}=-\frac{\lambda-\lambda_0}{\tau}+\Delta\sqrt{\frac{2}{\tau}}\xi(t),
\end{equation}
where $\xi$ is Gaussian white noise, $<\xi(t)>=0$, and $<\xi(t),\xi(t')>=\delta(t-t')$.

\begin{figure}[t]
\includegraphics[width=70mm]{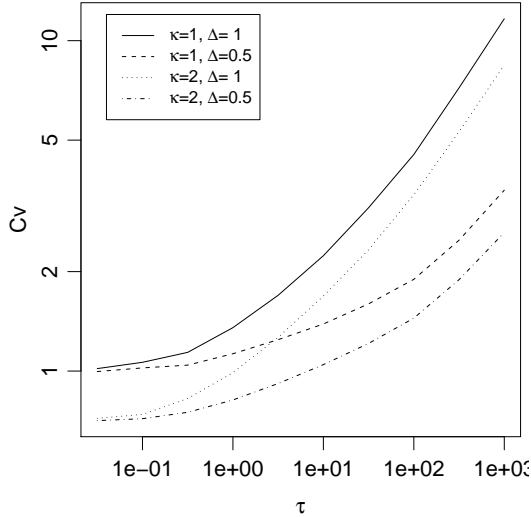}
 \caption{\label{TDGcv}$C_V$ for doubly stochastic gamma process with various 
values of time constant $\tau$ and rate amplitude $\Delta$.}
\end{figure}

Fig.~\ref{TDGlv} and Fig.~\ref{TDGcv} show $L_V$ and $C_V$ with $\lambda_0=1$ 
for various values of time constant $\tau$ and rate amplitude $\Delta$.
For simplicity, we consider sufficiently long spike sequences and 
assume that the values of $L_V$ and $C_V$ converge.
Fig.~\ref{TDGlv} shows that in the limit of a large time constant, the values
of $L_V$ converge to the value for the stationary case.
This means that the value of $L_V$ does not depend on the amplitude of the 
firing rate and has one-to-one correspondence with $\kappa$ in this limit.
Fig.~\ref{TDGcv} shows that $C_V$ depends on both $\kappa$ and $\Delta$ and
does not have one-to-one correspondence with $\kappa$.
Therefore, $L_V$ is better than $C_V$ for discriminating the intrinsic 
randomness of spike sequences.

This attractive property seems to stem from the fact that $L_V$ is the sum of
the dimensionless terms of consecutive interspike intervals.
By ``dimensionless'' we mean that the numerator and denominator have the same
dimension.
Every term in $L_V$ is normalized locally by the average of two consecutive
interspike intervals instead of the global average.
Intuitively, because the firing rates for two consecutive interspike intervals
can be regarded as the same in the slow limit, terms should be the same as
those for the stationary case.
On the other hand, $C_V$ is the variance around the global mean of the ISIs
and can be large for both the case where the firing rate fluctuates 
significantly and the case where the intrinsic randomness is large.
Therefore, we cannot distinguish the two cases based on the value of $C_V$.

\section{\label{sec4}Measure of local variation}
We extend $L_V$ without losing its attractive property described in the
previous section and find a better measure of intrinsic randomness.
We do this by focusing on the ISI statistics and imposing three symmetry
conditions: (1) time translation invariance, (2) time-scale transformation
invariance, and (3) time inversion invariance.

We assume the randomness of a spike train is constant over time and 
define the extended $L_V$ as
\begin{equation}
\widetilde{L_V} = \frac{1}{n-1}\sum_{i=1}^{n-1} f(T_i,T_{i+1}),
\end{equation}
where $T_1,T_2,...T_n$ are the observed ISIs and $f(x,y)$ does not depend on 
$i$ explicitly.
This form guarantees invariance under time translation ($i\rightarrow i+1$)
if $n$ is infinite.
Next, we assume that $f$ is invariant under the time-scale 
transformation ( $T\rightarrow kT$).
This requires that the denominator and numerator of $f$ have the same
dimension.
For simplicity, we assume that the dimension is two, so $f$ can be written as
\begin{equation}
f(x,y) = \frac{c_1 x^2 + c_2 xy + c_3 y^2}{c_4 x^2 + c_5 xy + c_6 y^2},
\end{equation}
which includes the original $L_V$ as a specific case.
In addition, because we do not distinguish increases from decreases in the
firing rate in terms of randomness, we impose time inversion invariance and
require 
\begin{equation}
f(x,y) = f(y,x).
\end{equation}
Thus, $f$ can be written as
\begin{equation}
f(x,y) = \frac{c_1 x^2 + c_2 xy + c_1 y^2}{c_4 x^2 + c_5 xy + c_4 y^2}.
\end{equation}
Note that the absolute value of $L_V$ does not matter in discriminant
analysis, and we can add (or multiply by) a constant to $f$.
Then, without loss of generality, $f$ can be written as
\begin{equation}
f(x,y) = \frac{xy}{x^2 + c_5 xy + y^2}.
\end{equation}
In addition, we can rewrite the denominator using $c=c_5+2$:
\begin{equation}
f(x,y) = \frac{xy}{(x-y)^2 + c xy}.
\label{f}
\end{equation}
Because each term in the denominator is non-negative, the necessary and
sufficient condition that the denominator always be positive is $c>0$.

As a result, $\widetilde{L_V}$ can be written as
\begin{equation}
\widetilde{L_V}(c) = \frac{1}{n-1} \sum_{i=1}^{n-1} \frac{T_i T_{i+1}}{(T_i-T_{i+1})^2 + c T_i T_{i+1}}.
\end{equation}
Note that the original $L_V$ corresponds to the case of $c=4$.
In this way, the measures satisfying the symmetries have only one degree of
freedom and can be parametrized by a single parameter.

\begin{figure}[t]
\includegraphics[width=70mm]{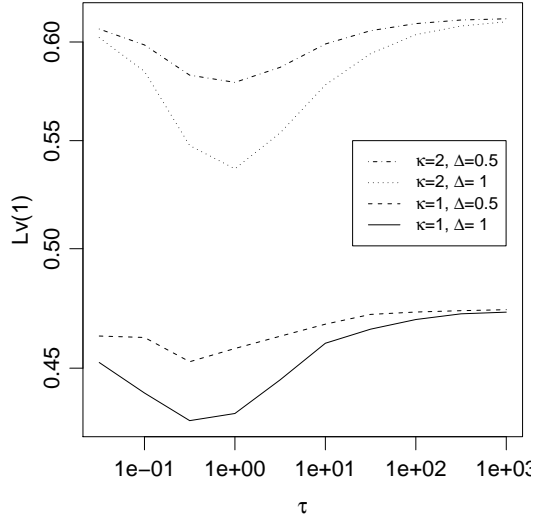}
 \caption{\label{lv1}$\widetilde{L_V}(1)$ for doubly stochastic gamma process
with various values of time constant $\tau$ and rate amplitude $\Delta$.}
\end{figure}

The $\widetilde{L_V}$ should have one-to-one correspondence to $\kappa$
like $L_V$ because of its symmetries.
In fact, it has the same values as those for the stationary case in the
limit of a large time constant for the doubly stochastic gamma process.
Fig.~\ref{lv1} shows that $\widetilde{L_V}(1)$ is independent of the rate 
amplitude, $\Delta$, and is a function of $\kappa$ in the limit.
The results for other values of $c$, for instance $\widetilde{L_V}(16)$, 
remain the same.

Thus, $\widetilde{L_V}(c)$ has one-to-one correspondence with $\kappa$.
However, this is not sufficient to make it a good measure.
We previously have considered only spike sequences with infinite length.
However, in practical experimental situations, data sizes are limited, and
$\widetilde{L_V}(c)$ varies widely by trial around the mean.
Similarly, if spike sequences are generated using a gamma distribution,
$\widetilde{L_V}(c)$ varies by trial for the finite spike sequence.
In the discrimination of intrinsic randomness, roughly speaking, the smaller
the variance, the higher the hitrate.
Thus, we next search for an optimal value of parameter c, where the variance
is the smallest.

\section{\label{sec5}Mutual information maximization principle}
We use the mutual information maximization principle to determine an optimal
measure.
We assume that the firing rate is constant over time and spike sequences are 
generated by a gamma distribution, as shown in Sec.~\ref{sec2}.
As shown in Sec.~\ref{sec3}, $\widetilde{L_V}$ does not depend on $\mu$.
Here we set $\mu=1$.
We consider the stationary case because it is tractable and
can be regarded as the slow change limit of the firing rate.
We show in Sec.~\ref{TD} that the optimal value of $c$ for the nonstationary
case does not differ significantly from that for the stationary case.

The optimal parameter value is determined so as to maximize the mutual
information between the coefficients and randomness.
Here we assume that a spike train consists of 100 ISIs because this is the 
typical length available from laboratory experiments.
$\widetilde{L_V}$ can be computed for a spike train, and the value of 
$\widetilde{L_V}$ varies among spike trains.
Even if spike trains are generated from the same distribution, the values of 
$\widetilde{L_V}$ can differ because the length of a spike train is finite.
As a result, the distribution of $\widetilde{L_V}$ can be obtained for one 
parameter set of the gamma distribution.
Thus, two distributions can be obtained from two types of spike trains.
The mutual information can be computed from the two distributions.
The bigger the mutual information, the better randomness ($\kappa$) can be
discriminated based on the observed $\widetilde{L_V}$.

Mutual information is calculated as follows.
Spike trains are generated from two gamma distributions with equal probability,
$\frac{1}{2}$.
The two distribution have different $\kappa$.
All the ISIs in a spike train are generated by using the same distribution.
We denote the distribution of $\widetilde{L_V}$ generated from the 
$i(=1,2)$-th gamma distribution as $p(x|i)$;
$p(x)(=\frac{1}{2}p(x|1)+\frac{1}{2}p(x|2))$ represents the distribution of 
$\widetilde{L_V}$ with no distinction of the source.
The entropy is defined as
\begin{equation}
H = -\int p(x) \ln p(x) dx.
\end{equation}
The noise entropy is defined as
\begin{equation}
H_{n}=-\frac{1}{2}\int p(x|1)\ln p(x|1) dx -\frac{1}{2}\int p(x|2)\ln p(x|2)dx.
\end{equation}
The mutual information is the difference,
\begin{equation}
I_m = H - H_{n}.
\end{equation}

The mutual information is the reduction in uncertainty about the spike trains
due to the knowledge of $\widetilde{L_V}$.
Mutual information is $0$ if two distributions of $\widetilde{L_V}$ are 
identical so that they cannot be distinguished .
Mutual information is $1$ if two distributions of $\widetilde{L_V}$ have no 
overlap, and only one sample of $\widetilde{L_V}$ is needed to distinguish 
them.

In the next section, we will show the results of a Monte Carlo simulation.
We calculated mutual information as a function of $c$ for various sets of 
randomness, $\kappa_1$ and $\kappa_2$.

\begin{figure}[t]
  \includegraphics[width=70mm]{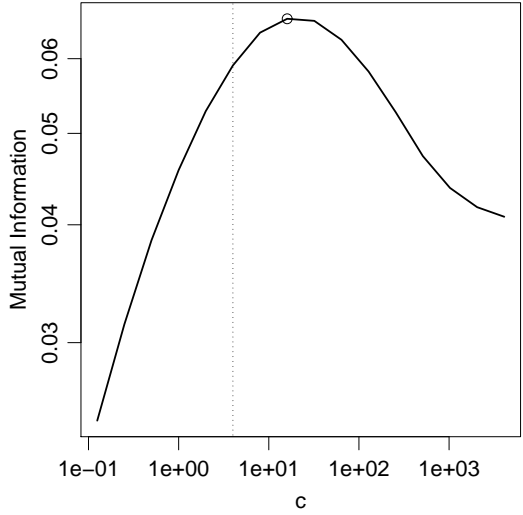}
  \caption{\label{minfo1}Mutual information with $\kappa_1=1,\kappa_2=1.1$. Open circle denotes peak. Dotted line is for $c=4$ corresponding to original $L_V$. Mutual information has a peak with $c$ larger than $4$.}
\end{figure}

\section{\label{sec6}Results}
Fig.~\ref{minfo1} shows the mutual information with $\kappa_1=1$ and 
$\kappa_2=1.1$; $\kappa_1$ and $\kappa_2$ are the shape parameters of two
gamma distributions and $c$ is the parameter in $\widetilde{L_V}(c)$.
We set the number of ISIs per spike train, n, to 100.
The mutual information has a peak, whose location we denote by $c_{peak}$.
The vertical line represents $c=4$, which corresponds to the original $L_V$.
Since $c_{peak} (\approx16)$ is bigger,
the optimal coefficient in this case is not the original $L_V$ but
$\widetilde{L_V}(16)$.
However, $c_{peak}$ depends on various parameters, and we will examine how it 
depends on the number of ISIs per spike train, $\kappa_1$ and $\kappa_2$, in
what follows.
We can use the maximum likelihood estimator of $\kappa$ as a measure instead 
of $L_V$, and the peak value of the mutual information for $\kappa$ is 0.097.
(For the maximum likelihood estimator, see Appendix \ref{appendixB}.)
The peak value for $L_V$ is about 0.066, which is smaller than that for the 
maximum likelihood estimator.
We nonetheless use $L_V$ because the maximum likelihood estimator cannot be
applied to the nonstationary case.
In the cases where the firing rate is time-dependent, the mutual information
for $L_V$ can be much higher than that for the maximum likelihood estimator,
as we will show in Sec.~\ref{TD}.

\begin{figure}[t]
  \includegraphics[width=70mm]{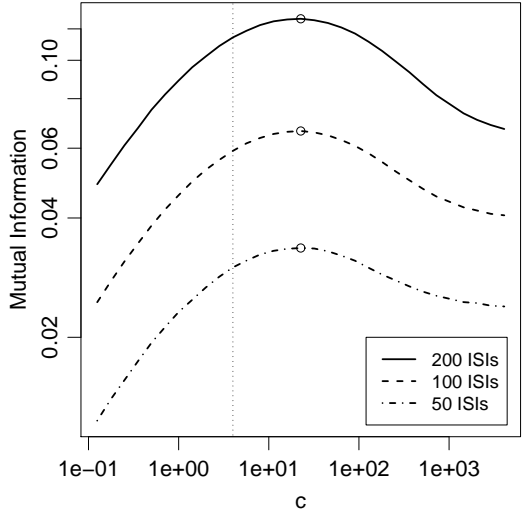}
  \caption{\label{minfo-nspike}Mutual information for various numbers of ISIs
per spike sequence with $\kappa_1=1,\kappa_2=1.1$.
Open circles denote peaks.
Dotted line is for $c=4$ corresponding to original $L_V$.
Peak location almost does not depend on number of ISIs.}
\end{figure}

Fig.~\ref{minfo-nspike} shows the mutual information for various numbers of
ISIs per spike train.
While the mutual information increases with the number of ISIs,
the peak location remains almost the same.
Although we show only the case for $\kappa_1=1,\kappa_2=1.1$,
the other cases have similar results.
Therefore, we set the number of ISIs per spike train to $100$.

\begin{figure}[t]
  \includegraphics[width=70mm]{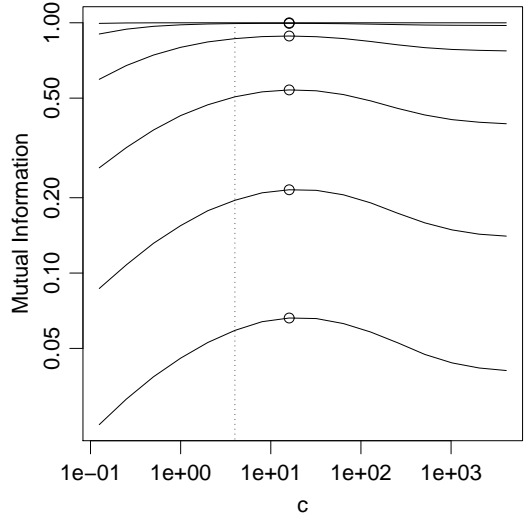}
  \caption{\label{minfo-0to1}Mutual information for various $\kappa_2$ with $\kappa_1=1$.
Lines are for $\kappa_2=0.1, 0.2, 0.4, 0.8, 1.6$ and $3.2$ from below.
Open circles denote peaks.
Dotted line is for $c=4$ corresponding to original $L_V$.
Peak location almost does not depend on $\kappa_2$.}
\end{figure}

Fig.~\ref{minfo-0to1} shows the mutual information with $\kappa_1=1$ and 
various $\kappa_2$.
As $d\kappa(=\kappa_2-\kappa_1)$ increases, the mutual information approaches
$1$.
The peak location remains almost unchanged $(c_{peak}\approx16)$.
For $\kappa_2=3.2$, the mutual information is almost $1$, and 
the two distributions are completely distinguishable.
In general, $c_{peak}$ largely depends on $\kappa_1$ and is almost independent
of $\kappa_2$.

\begin{figure}[t]
  \includegraphics[width=70mm]{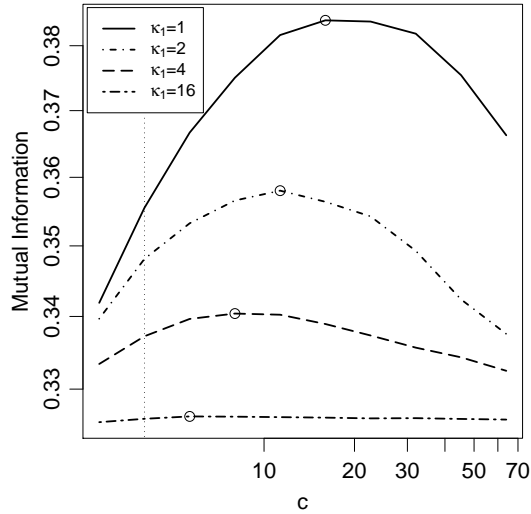}
  \caption{\label{minfo-peak}Mutual information for various $\kappa_1$ with $\kappa_2=1.3\kappa_1$.
Open circles denote peaks.
Dotted line is for $c=4$ corresponding to original $L_V$.
Peak location decreases as $\kappa_1$ increases.}
\end{figure}

Fig.~\ref{minfo-peak} shows the mutual information with $\kappa_2=1.3\kappa_1$
and various $\kappa_1$.
The peak location decreases with increasing $\kappa_1$.
For $\kappa_1=16$, the original $L_V$ is nearly optimal
($c_{peak}\approx4\sqrt{2}$).
Since reported experimental data can be well fitted by a gamma distribution
with $\kappa\approx16$ \cite{baker}, $L_V$ seems to be optimal not for the
Poisson data but for the experimental data.

\section{\label{sec7}theoretical analysis}
In this section we analyze the property of the mutual information
theoretically.
For simplicity, we do two approximations.

First, we consider the limit of a large number of ISIs per spike train and 
approximate the distribution of $L_V$ by using the normal distribution.
Although this approximation is not good for $c\approx 0$,
the peak location is far larger than 0 and can be discussed within this 
approximation.

In addition, we consider the limit of small $d\kappa$.
In the limit, the mutual information can be written using the Fisher
information \cite{lehmann} as
\begin{equation}
I_m = \frac{1}{8}J(p(x,\kappa))d\kappa^2,
\label{eighth}
\end{equation}
where the Fisher information is defined as
\begin{equation}
J(p(x,\kappa))= Ex((\frac{d\log p(x,\kappa)}{d\kappa})^2).
\end{equation}
This relation can be easily derived.
We represent two $L_V$ distributions as
\begin{equation}
p_1(x)=\frac{1}{\sqrt{2\pi\sigma(\kappa)^2}}e^{-(x-m(\kappa))^2/2\sigma(\kappa)^2}
\end{equation}
and
\begin{equation}
p_2(x)=\frac{1}{\sqrt{2\pi\sigma(\kappa+d\kappa)^2}}e^{-(x-m(\kappa+d\kappa))^2/2\sigma(\kappa+d\kappa)^2} .
\end{equation}
Inserting these equations into the definition of the mutual information and 
expanding by $d\kappa$ to the second order lead to the relation.

The Fisher information can be explicitly written as
\begin{equation}
J=\frac{m'(\kappa)^2+2\sigma'(\kappa)^2}{\sigma(\kappa)^2}.
\end{equation}
Because $\sigma^2$ is inversely proportional to $N$, $\sigma$ can be written as
\begin{equation}
\sigma = \frac{\sigma_0}{\sqrt{N}}.
\end{equation}
The Fisher information can then be approximated as
\begin{eqnarray}
J/N &=& \frac{m'(\kappa)^2+2\frac{1}{N}\sigma_0'(\kappa)^2}{\frac{1}{N}\sigma_0(\kappa)^2}\frac{1}{N}\nonumber\\
  &\simeq& \frac{m'(\kappa)^2}{\sigma_0(\kappa)^2},
\end{eqnarray}
where $m'$ and $\sigma_0$ depend on only $\kappa$ and $c$.
As a result, the mutual information can be written as
\begin{equation}
I_m = \frac{1}{8} \frac{m'(\kappa,c)^2}{\sigma_0(\kappa,c)^2} N d\kappa^2.
\end{equation}

Thus, $I_m$ is proportional to $N$ and $d\kappa^2$.
The $c$ dependency of $I_m$ stems from only $m'$ and $\sigma_0$.
Therefore, when $N$ or $d\kappa$ changes, the absolute value of the mutual
information changes while the peak location does not change.
This is consistent with our numerical results in which the peak location
did not depend on $N$ and $d\kappa$.
The peak location can be explained by an interplay of $m'$ and $\sigma_0$.
However, $m$ and $\sigma_0$ cannot be predicted solely by our theory.
Numerical calculations are necessary for finding the peak location.

\section{\label{TD}nonstationary case}
We considered the discrimination of randomness for the stationary gamma
process in the previous sections.
However, it has been reported that experimental data can be explicable by 
the rate-modulated gamma process \cite{baker}.
Therefore, we consider the rate-modulated gamma process in this section.
We show two simple cases in which the firing rate decreases monotonically
or changes stepwise.

\subsection{monotonically decreasing firing rate}

\begin{figure}[t]
  \includegraphics[width=70mm]{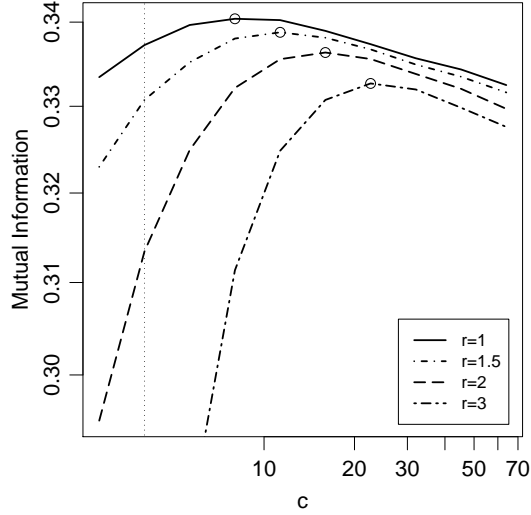}
  \caption{\label{td-4}Mutual information for monotonically decreasing firing rate for various $r$ with $\kappa_1=4$ and $\kappa_2=5.2$.}
\end{figure}

We consider a simple rate-modulated case and show that the peak location of 
the mutual information, $c_{peak}$, tends not to change if the firing rate
fluctuates significantly.
We generate the ISIs by again using a gamma distribution.
We assume that the mean ISI increases monotonically.
For simplicity, we set $\mu_i=r^i$, where $\mu_i$ denotes the mean of the
i-th ISI.
We simply align $n$ ISIs to make a single spike train as before.
The value of $\kappa$ does not change within the train.
The mutual information is calculated for two spike trains with different 
values of $\kappa$.

Fig.~\ref{td-4} shows the mutual information for $\kappa_1=4$ and
$\kappa_2=5.2$.
The peak location decreases gradually from the stationary value as r increases.
However, only extreme cases, in which the firing rates decrease more than 1.5
times one after another, are plotted.
For realistic cases, $c_{peak}$ changes only slightly.
For example, the ratio between the last and first mean ISI is
\begin{equation}
\frac{\mu_n}{\mu_1}=r^{n-1},
\end{equation}
and the ratio is 2.678033 for $r=1.01$ and $n=100$ and 12527.83 for
$r=1.1$ and $n=100$.
This illustrates that the $1.5$ used for r is extremely large.
Similar results were obtained for different values of $\kappa$, so
$c_{peak}$ apparently tends not to change even if the firing rate fluctuates.
This result is not restricted to the decreasing firing rate case.
For example, the mean $\widetilde{L_V}$ remains the same if a small and a 
large mean ISI appear alternately instead of the firing rate increasing
monotonically.
It thus appears that the peak location of the mutual information is almost
independent of the firing rate if the variation in the firing rate is small.

\subsection{stepwise changing firing rate}
\begin{figure}[t]
  \includegraphics[width=70mm]{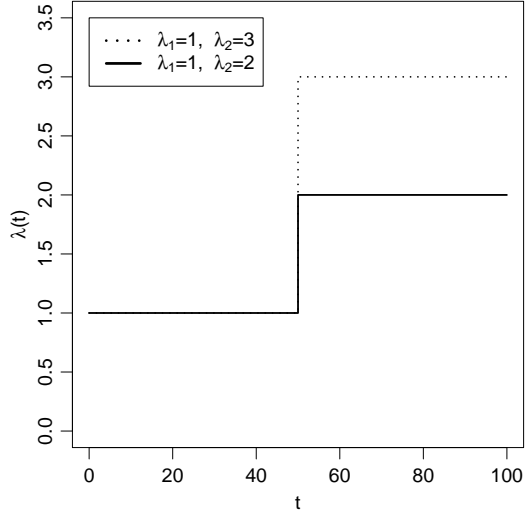}
  \caption{\label{stairs}
Schematic diagram of stepwise increasing firing rate. Firing rate shifts from $1$ to $\lambda_2$ at $t=50$.}
\end{figure}

So far we have considered only $\widetilde{L_V}(c)$.
However, the maximum likelihood estimator, $\hat{\kappa}$, should be better
for the stationary case.
Here we consider the case of a stepwise changing firing rate to show why
we favor $\widetilde{L_V}(c)$ nonetheless.
In a word, $\hat{\kappa}$ is not good for the nonstationary case because it is
the maximum likelihood estimator for the stationary case, as shown in Appendix
\ref{appendixB}.
In principle, the firing rate at every small time bin can be estimated for the
nonstationary case.
However, doing so requires many spike sequences and the firing rate profile
must be the same for all the sequences.
Therefore, it is not practical for many realistic cases.
Instead we consider simple measures like $\widetilde{L_V}(c)$ and
$\hat{\kappa}$ even in the nonstationary case.
In this section, we compare $L_V$ and $\hat{\kappa}$ for the nonstationary
case.

\begin{figure}[t]
  \includegraphics[width=70mm]{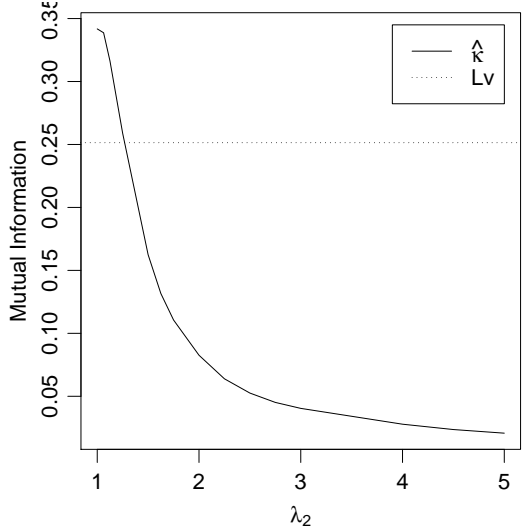}
  \caption{\label{step-mle}Mutual information for stepwise increasing firing rate with $\kappa_1=16$ and $\kappa_2=20$.}
\end{figure}

Consider the case in which the firing rate is stepwise increasing, as shown in
the Fig.~\ref{stairs}.
At time $t=50$, it shifts from $1$ to $\lambda_2$.
Two types of spike trains, with $\kappa_1=16$ and $\kappa_2=20$, are generated
based on the firing rate profile.
Fig.~\ref{step-mle} shows the mutual information for these trains
when $L_V$ or $\hat{\kappa}$ is used as a measure.
The mutual information for $L_V$ is independent of $\lambda_2$ in the limit of
a large number of ISIs per train.
The reason is that $L_V$ is independent of the firing rate for the stationary
case and in this case the firing rate is constant over time except for the
discontinuous point.
The contribution of the term in $L_V$ that cross the discontinuous point is
$O(1/n)$ and is small if the number of ISIs is large enough.
We plotted the value for the stationary case, neglecting the contribution for
simplicity.
On the other hand, the mutual information for $\hat{\kappa}$ decreases as
$\lambda_2$ increases.
For example, when the firing rate increases 1.5 times, the mutual information 
for $L_V$ is larger than that for $\hat{\kappa}$.

Thus, for a stepwise increasing firing rate, $L_V$ is better than
$\hat{\kappa}$.
This type of sudden change can be observed when a visual stimulus is presented
to a monkey at a given time.
The result remains almost the same for the stepwise firing rate with multiple 
discontinuous points in the limit of a large number of ISIs.
In addition, $\hat{\kappa}$ depends on both $\kappa$ and the amplitude of the
firing rate, as shown in Sec.~\ref{sec2} for $C_V$.
Therefore, $L_V$ is a better measure of intrinsic randomness.

\section{summary and discussion}
In this study, we sought a measure more effective than the local variation of interspike 
intervals, $L_V$, in discriminating spike trains based on the degree of intrinsic spiking 
irregularity. 

We first compared characteristics of the conventional coefficient of variation, $C_V$, 
and the local variation, $L_V$. The coefficient of variation, $C_V$, measures a global 
variability of ISIs, and therefore depends on not only the local irregularity of ISIs but 
also the rate fluctuation, which would naturally manifest itself in \textit{in vivo} 
neuronal spiking conditions. In contrast, the local variation, $L_V$, measures only a 
stepwise variability of ISIs, and therefore does not depends significantly on a rate 
fluctuation. It was revealed that $L_V$ is superior to $C_V$ in detecting some intrinsic 
spiking irregularity specific to individual neurons \textit{in vivo} 
\cite{shinomoto,shinomoto7}.

For a spike train of a finite number of ISIs derived from a given point process, the value 
of $L_V$ as well as $C_V$ varies from trial to trial. The goodness of a 
coefficient is quantified by its narrow distribution of values among 
spike trains derived from the same point process and 
the small overlap of this distribution with the distribution obtained 
from spike trains derived from a different point process. In other words,
we sought a new coefficient that maximizes the mutual information between 
spike sequences created from different renewal gamma processes.

For this purpose, we adopted a rational expression of quadratic functions of 
consecutive interspike intervals that is the same form as $L_V$, and 
searched for the optimal parameter of the coefficient.  The optimal parameter of the 
coefficient depends on the choice of the point processes that are to be 
discriminated. It was found that the original $L_V$ is not optimal for near random 
(Poisson) point processes, but is optimal for more regular spike trains. In this way, if we 
have preliminary knowledge of the spiking irregularities of the point processes, 
we are able to propose a better coefficient than the original $L_V$ for the purpose of 
discriminating spike trains.

We generated spike sequences entirely by using a stationary or rate-modulated gamma process.
The reason is as follows.
The Poisson process, in which the firing rate is represented as a function of time from
stimulus onset, is widely used in spike data analysis \cite{richmond}.
However, the statistical properties of spike sequences cannot be fully
captured by the rate-modulated Poisson process \cite{berry,reich,keat,pillow}.
In other words, spike probability depends on the past spike times due to the 
so-called refractory period.
A gamma process is a Poisson process with an additional parameter representing
a kind of refractory period.
Baker et al. showed that the spike pattern recorded from primary and 
supplementary motor areas is explicable by a gamma process \cite{baker}.

We considered only mutual information as a measure for discriminating
two spike trains.
However, the Kullback-Leibler divergence $D(p_1,p_2)$ is a well-known measure
of the dissimilarity of two distributions, too.
It is also proportional to the Fisher information,
$D=\frac{1}{2}J(p(x,\kappa))d\kappa^2$, 
under the same approximation as described in Sec.~\ref{sec7}.
Note that the coefficient is $\frac{1}{2}$ instead of $\frac{1}{8}$, as seen in
Eq.~(\ref{eighth}), for the mutual information.
However the coefficient is irrelevant to the peak location.
Thus, the Kullback-Leibler divergence leads to the same results as mutual
information.
Nontheless, we used mutual information because it is symmetrical in terms of
two distributions.
The Kullback-Leibler divergence is not symmetrical.
Its value changes if the two distributions are interchanged.
The Kullback-Leibler divergence becomes symmetrical in the limit of a small
difference of two distributions, where it is proportional to the Fisher
information.

In previous studies, various measures were computed for mathematical models
\cite{lansky,feng,shinomoto3}.
However, the focus was only on the expectations for the measures.
In discrimination tasks, the variance of a measure is more important than the expectation.
For example, consider the case in which the expectations of a measure for two
different types of spike sequences differ considerably.
If the variances are very large, discriminating the two sequences is difficult.
In addition, if the definition of a measure is changed, for example,
multiplied or added to by a constant, the expectation changes, but the mutual
information never changes.
Therefore, in this paper we focused on the variance and searched for the
measure that maximizes the mutual information.

\appendix
\section{\label{appendixA}Orthogonal coordinates for gamma distribution}
We show that $\kappa$ and $\mu$ are orthogonal coordinates in the sense of 
information geometry.
The theory of information geometry is described elsewhere \cite{amari2,amari3},
and there are applications to neuroscience \cite{tatsuno,tatsuno2,nakahara}.

For the purpose of proving the orthogonality, it suffices to demonstrate that
the Fisher information matrix is diagonal.
The Fisher information matrix is defined as
\begin{equation}
g_{ij} = \int^{\infty}_{0}\frac{\partial \log p(T)}{\partial \xi^i}\frac{\partial \log p(T)}{\partial \xi^j}p(T)dT,
\end{equation}
where $\xi^1=\mu$ and $\xi^2=\kappa$.
The log-likelihood can be written as
\begin{equation}
\log p(T) = \kappa \log(\frac{\kappa}{\mu}) + (\kappa -1) \log T - \log \Gamma(\kappa) - \frac{T \kappa}{\mu}.
\end{equation}
The derivatives of the log-likelihood are
\begin{equation}
\frac{\partial \log p(T)}{\partial \mu} = - \frac{\kappa}{\mu} + \frac{T \kappa}{\mu ^2}
\end{equation}
and
\begin{equation}
\frac{\partial \log p(T)}{\partial \kappa} = \log \frac{\kappa}{\mu} +1 + \log T - \psi (\kappa) - \frac{T}{\mu},
\end{equation}
where $\psi(\kappa)=(\log\Gamma(\kappa))'$.
The matrix elements can be written as 
\begin{equation}
g_{\mu\mu}=\frac{\kappa}{\mu ^2},
\end{equation}
\begin{equation}
g_{\mu\kappa}=g_{\kappa\mu}=0,
\end{equation}
\begin{equation}
g_{\kappa\kappa}=\psi(\kappa)' - \frac{1}{\kappa}.
\end{equation}
Thus, the Fisher information matrix is diagonal at every point.
According to the theory of information geometry, it is always possible to 
choose orthogonal coordinates for an exponential family of distributions that
includes the gamma distribution as a specific case.

The Fisher information matrix has the meanings described below.
When $\mu$ and $\kappa$ are estimated from a finite number of samples,
the estimated values are not necessarily the same as the true value.
The value of the maximum likelihood estimator varies depending on the sample
sets, and its variation around the true value can be approximated by a normal 
distribution whose variance is the inverse of the Fisher matrix if the sample
size is sufficiently large \cite{lehmann}.
Thus, the diagonality of the Fisher matrix means that the variations in the 
maximum likelihood estimators of $\mu$ and $\kappa$ are uncorrelated.

\section{\label{appendixB}Maximum likelihood estimation for gamma distribution}
Let $T_1,T_2,...,T_n$ be observed ISIs.
We would like to estimate the true values of $\mu$ and $\kappa$ from them.
The log-likelihood is defined as
\begin{equation}
l \equiv \ln(p(T_1)p(T_2)...p(T_n))
\end{equation}
and can be written as
\begin{equation}
l = n \kappa \ln \frac{\kappa}{\mu} - n \Gamma(\kappa) + (\kappa -1) \sum \ln T_i - \frac{\kappa}{\mu} \sum T_i .
\end{equation}
The maximum likelihood estimators must satisfy both 
$\frac{\partial l}{\partial \mu} = 0$ and 
$\frac{\partial l}{\partial \kappa} = 0$.
The derivatives of the log-likelihood are
\begin{equation}
\frac{\partial  l}{\partial \mu} = \frac{\kappa}{\mu^2}\sum T_i - n \frac{\kappa}{\mu}
\end{equation}
and
\begin{equation}
\frac{\partial l}{\partial \kappa} = \sum \ln T_i - \frac{1}{\mu}\sum T_i + n \ln\frac{\kappa}{\mu} + n -n\psi(\kappa).
\end{equation}
Then, $\hat{\mu}$ can be explicitly obtained as
\begin{equation}
\hat{\mu} = \frac{1}{n}\sum T_i,
\end{equation}
and $\hat{\kappa}$ must satisfy
\begin{equation}
\frac{1}{n}\sum \ln T_i - \ln\frac{1}{n}\sum T_i = \psi(\hat{\kappa})-\ln\hat{\kappa},
\label{mle}
\end{equation}
where $\psi(\hat{\kappa})=(\log\Gamma(\hat{\kappa}))'$.
This equation cannot be solved explicitly for $\hat{\kappa}$.
However, the right side of the equation is a monotonic function of 
$\hat{\kappa}$, and we can obtain $\hat{\kappa}$ by numerical iteration.

Instead of a lengthy numerical iteration, we can use the moment estimator.
According to Eq.~(\ref{expectation}), we can estimate the true $\kappa$ from
the sample mean and variance:
\begin{equation}
\kappa=\frac{Ex(T)^2}{Var(T)}.
\label{cv}
\end{equation}
In fact, the right side of Eq.~(\ref{cv}) can be rewritten as 
$\frac{1}{C_V^2}$.
Thus, we can regard $C_V$ as a moment estimator.
However, the moment estimator is worse than the maximum likelihood estimator,
especially when $\kappa$ is close to $1$ \cite{cox}.
Nevertheless, it is good as a first approximation, and we can use it as the
initial value of the numerical iteration in maximum likelihood estimation.

Another way to avoid numerical iteration is to use the left side of
Eq.~(\ref{mle}) as a measure.
In discriminant analysis, we do not need to estimate $\kappa$ because 
the left side of Eq.~(\ref{mle}) has one-to-one correspondence with
$\hat{\kappa}$ and has the same information as $\hat{\kappa}$.



\bibliography{pre}

\end{document}